\pdfoutput=1
\documentclass[lettersize,journal]{IEEEtran}
\usepackage{amsmath,amsfonts}
\usepackage{array}
\usepackage{textcomp}
\usepackage{stfloats}
\usepackage{url}
\usepackage{verbatim}
\usepackage{graphicx}
\def\BibTeX{{\rm B\kern-.05em{\sc i\kern-.025em b}\kern-.08em
    T\kern-.1667em\lower.7ex\hbox{E}\kern-.125emX}}
\usepackage{balance}

\usepackage{amssymb}
\usepackage{lscape}
\usepackage{afterpage}
\usepackage{bm}    
\usepackage{float}
\usepackage{placeins}
\usepackage{enumitem}
\usepackage{listings}
\usepackage{multirow}
\usepackage{tikz}
\usetikzlibrary{decorations.pathreplacing}
\usepackage[normalem]{ulem}
\usepackage{ragged2e}
\renewcommand{\arraystretch}{1}  
\usepackage{pifont}
\usepackage{subcaption}
\usepackage{upgreek}
\usepackage{relsize}
\usepackage{microtype}
\usepackage{algpseudocode}
\usepackage{makeidx}  
\usepackage{geometry}
\usepackage{setspace}
\usepackage{color}
\usepackage[utf8]{inputenc}
\usepackage{wrapfig}
\usepackage[font={small}]{caption}
\usepackage[labelfont=bf]{caption}
\usepackage[noadjust]{cite}
\usepackage[square,numbers]{natbib}
\usepackage{cite}
\usepackage{pdflscape}
\usepackage{etoolbox}
\usepackage{alltt}
\usepackage{framed}
\usepackage{cleveref}
\usepackage[thinlines]{easytable}
\usepackage[linesnumbered,ruled,vlined]{algorithm2e}

\SetCommentSty{mycommfont}
\SetKwInput{KwInput}{Input}               
\SetKwInput{KwOutput}{Output}             
\begin{document}

\title{A Novel Information Workflow for Structural Behavioural Analysis in Dynamic Attributed Graphs}
\author{
Gatadi Ashwitha,
K.~Swarupa~Rani\thanks{Corresponding author: K. Swarupa Rani (swarupacs@uohyd.ac.in).}%
\thanks{The authors are with the School of Computer and Information Sciences,
University of Hyderabad, CR Rao Road, Gachibowli, Hyderabad, Telangana 500046, India
(e-mail: 20MCPC14@uohyd.ac.in; swarupacs@uohyd.ac.in).}
}
\maketitle
\begin{abstract} 
Social systems play a vital role through interactions around shared interests and the maintenance of relationships, which are part of society. These can be represented as networks, and their analysis plays a crucial role in dynamic environments, which can be achieved through link prediction and community detection tasks. Several link prediction approaches estimate new relationships, such as suggesting friends on social media. On the other hand, a social faction is a group of connected people that reveals the internal structure of the social system and can be identified through community detection approaches. The evolution of social systems can be understood using these approaches by analysing their structure and behaviour. To handle this, a combination of these approaches, along with interaction-based analysis, is used to form cohesive communities, thereby enabling the study of social systems dynamically. In addition, attribute information plays a critical role. Most real-world graphs, like Facebook and Twitter, have attribute information that provides context, and integrating them with structural information offers deeper insights into patterns, which is a critical task. However, existing approaches have limitations in handling dynamics of social systems. Additionally, no existing approaches in the literature help identify potential links and form densely connected communities for attributed graphs. To address these challenges, we have proposed an information workflow, i.e., $inc$-$LPCD_{AG}$ (\underline{inc}remental \underline{L}ink \underline{P}rediction and \underline{C}ommunity \underline{D}etection in \underline{A}ttributed \underline{G}raphs). We conducted experiments to evaluate efficiency, and results demonstrate the effectiveness of our proposed workflow by analysing and understanding structural and behavioural processes.
\end{abstract}
\section{Introduction}\label{intro}
Graphs consist of nodes (vertices), which are individuals, and edges (links), are the relationships between these individuals \cite{ModifiedLPA}. Different applications, such as web graphs, social networks \cite{IEEE2}, \cite{IEEE1}, and biological networks \cite{CSADW}, can be represented as complex graphs, containing large number of nodes with relationships following some pattern, which will help in predicting the links, identifying the spreader nodes, forming communities, and searching for the maximal influence nodes \cite{CSADW}.

In these real-world graphs, nodes are often associated with attributes, along with structural information \cite{CSADW}, forming attributed graphs that provide detailed information to help analyse tasks more accurately. Also, most real-world graphs are sparse, with fewer connections among nodes. Therefore, combining node attributes and structural information in these sparse graphs provides more meaningful information.

One of the most critical tasks in complex graphs is Link Prediction, which predicts potential links from the given graph structure. Some applications include movie recommendation \cite{MovieRecommendation} and friend recommendation \cite{FriendRecommendation}. Traditional similarity-based approaches \cite{LINE} for link prediction use only the structural information of the graph to predict links. When applied to attributed graphs, these approaches may result in predicting the maximum number of links between users with different characteristics rather than those with similar characteristics \cite{Sentiment}, as they ignore node attribute information. Therefore, combining node attributes and structural information for link prediction has become an important research direction. Several approaches have been proposed in the literature, such as CSADW \cite{CSADW}, CPAGCN \cite{CPAGCN}, RGNMF-AN \cite{RGNMF-AN}, and PaGNN \cite{PaGNN} that combines both node attributes and structural information of the graph.

Another significant task in complex graphs is Community Detection, which involves identifying the groups of nodes that are more densely connected \cite{ModifiedLPA}. According to Eric et al. \cite{LINE}, different community detection algorithms, including Girvan-Newman, Louvain, Label Propagation, WalkTrap, and EigenVector, utilize the structural information of the graph to detect communities. However, identifying communities based solely on structural information may not be efficient when the graph contains node attributes. Therefore, there is a need for a more comprehensive approach that combines node attributes and structural information of the graph to provide a deeper understanding of complex graphs \cite{ModifiedLPA}. Several existing approaches, such as CoDeDANet \cite{CoDeDANet}, ASMsg \cite{ASMsg}, and AGGMMR \cite{AGGMMR}, effectively use both node attributes and structural information of the graph for community detection.

Although link prediction and community detection are studied as separate tasks, they are closely related in real-world networks. For this reason, some approaches combine both tasks. There are approaches in the literature for link prediction using community structure information, which help predict links more accurately \cite{Sentiment}, assuming that links within communities are more likely to be predicted than those between communities \cite{LPDN}, \cite{CPAGCN}.

On the other hand, few approaches exist for community detection using link prediction to improve the quality of communities. As most available graphs may contain incomplete information such as missing links, detecting communities in these graphs may not be efficient \cite{LINE}. To accomplish this task, the LINE \cite{LINE} approach predicts the links before detecting the communities. It is a static approach, and when used in dynamic scenarios, whenever links are predicted, the communities are detected from scratch, which is computationally expensive. One of our works, the LPCD \cite{LPCD}, is an incremental approach that predicts links and detects communities in a dynamic environment using structural information of the graph. When applied to attributed graphs, it ignores node attributes. Hence, we enhanced this approach to the graphs associated with node attributes and proposed $inc$-$LPCD_{AG}$ information workflow (\underline{\textbf{in}}cremental \underline{\textbf{L}}ink \underline{\textbf{P}}rediction and \underline{\textbf{C}}ommunity \underline{\textbf{D}}etection for \underline{\textbf{A}}ttributed \underline{\textbf{G}}raphs)

\textbf{Problem Statement}
Given an attributed graph G$=$(V, E, A) where `V' represents the set of nodes, `E' represents the set of links, and `A' represents the node attributes, the objective is to predict links $LP = \{LP_1, LP_2, \dots , LP_i\}$, where each $LP_i$ represents the predicted link. These predicted links are added to the graph. Subsequently, the community structure CD $=$ \{$CD_1, CD_2, \dots, CD_i $\}, where each $CD_i$ represents a community in the graph, is updated to reflect the changes caused by adding the new links. 

The main contribution of our proposed information workflow includes:
\begin{itemize}
	\item Proposed an information workflow with two phases for link prediction and community detection in dynamic attributed graphs ($inc$-$LPCD_{AG}$).
	\item Designed a Dynamic CSADW algorithm for link prediction in Phase-I.
	\item Developed an efficient community detection algorithm in Phase-II and named it as Dynamic I-Louvain.	
\end{itemize}
The rest of the paper is organized as follows: Section 2 reviews existing link prediction and community detection approaches. Section 3 details our proposed information workflow. Section 4 presents the experimental results. Section 5 provides the conclusions and outlines possible directions for future work.

\section{Related Work}\label{relatedwork}
There are different approaches in the literature for link prediction and community detection in both non-attributed and attributed graphs. This section explores different approaches that exist for link prediction \cite{FSFDW}, \cite{CSADW}, and community detection \cite{ModifiedLPA}, \cite{TANMF} in an attributed graph. Also, a combination of both in the literature \cite{LPDN}, \cite{CLP-ID}, \cite{LINE}, \cite{LPCD} that are applied to non-attributed graphs.

\subsection{Approaches for Link Prediction}
Link prediction in attributed graphs mainly depends on node attributes and structural information of the graph. The CSADW \cite{CSADW} and FSFDW \cite{FSFDW} approaches predict links by integrating structural and node attribute information using the Deepwalk methods. PaGNN \cite{PaGNN} uses GNN for link prediction by combining edge-centric and node-centric information through novel aggregation and broadcasting operations introduced by Yang et al. (2021). RGNMF-AN \cite{RGNMF-AN} approach introduces the SARWS method to predict links by computing higher-order proximities, assuming that attribute information is more valuable in determining higher-order proximities. The CPAGCN \cite{CPAGCN} model, which consists of two layers, utilizes the AGCN layer to generate node embeddings by combining node attributes and structural information. These node embeddings are given to the MLP layer to compute scores representing the likelihood of forming links between node pairs. 

\subsection{Approaches for Community Detection}
There are different approaches in the literature for community detection in attributed graphs that detect communities using the node attributes and structural information of the graph. Lu et al. introduced the TANMF and TASNMF models \cite{TANMF}, which are parameter-free and use non-negative matrix factorization to detect communities by combining node attributes and structural information. I-Louvain \cite{ILouvain} algorithm incorporates inertia-based modularity to detect communities in attributed graphs. Malhotra et al. \cite{ModifiedLPA} introduced the Modified LPA method, addressing the issue of randomness in label selection by assigning a specific label when multiple labels with maximum value exist for a node. AGGMMR \cite{AGGMMR} approach begins by constructing an augmented graph that integrates attribute node information into the graph. It then uses modularity maximization and a weight learning mechanism to identify the communities. This approach works in a static environment. Therefore, Chen et al. \cite{incAGGMMR} enhanced the approach and proposed the inc-AGGMMR \cite{incAGGMMR}, which is suitable for detecting communities in dynamic environment. CoDeDANet \cite{CoDeDANet} detects communities in dynamic attributed graphs by using spectral clustering to integrate node attributes and structural information of the graph.
\subsection{Synergistic relationship between Link Prediction and Community Detection}
Some approaches \cite{LPDN}, \cite{CLP-ID} use community structure information of the graph to predict the links, assuming that the links are more likely to form within the communities. The CLP-ID \cite{CLP-ID} algorithm predict the links within the communities using the information of community structure and information diffusion. COMMLP \cite{LPDN}, predict links using machine learning models by generating a feature set from the structural and community-based information.

On the other hand, some approaches claim that the existing graphs often contain missing links, which can reduce the effectiveness of detected communities \cite{LINE}. Therefore, to address these issues, links are predicted before detecting communities. The LINE \cite{LINE} approach first predicts missing links, updates the graph, then detects communities, and finally compares the new communities with the initially formed communities to evaluate their quality. However, this is a static approach. One of our works, LPCD \cite{LPCD}, has extended the LINE \cite{LINE} approach to predict links and detect communities incrementally.  

There are different link prediction and community detection approaches in the literature; however, there is no unified approach that can handle attributed graphs, predict missing links and update the graph, and detect communities in a dynamic environment. Hence, there is a need for such a unified approach in real-world scenarios to handle attributed graphs in a dynamic environment. To address this, we have proposed \textbf{\underline{inc}}remental \textbf{\underline{L}}ink \textbf{\underline{P}}rediction and \textbf{\underline{C}}ommunity \textbf{\underline{D}}etection for \textbf{\underline{A}}ttributed \textbf{\underline{G}}raphs ($inc$-$LPCD_{AG}$) information workflow and discussed in Section \ref{approach}

\section{Link Prediction and Community Detection in Dynamic Graphs} \label{approach}
As discussed in Section \ref{relatedwork}, there are static approaches to predict the links either by using community structure or structural information of the graph. Some approaches \cite{LPDN}, \cite{CLP-ID} predict links using community structure information, resulting in a higher probability of predicting links within communities. These approaches may improve link prediction accuracy but limit their ability to capture potential links between communities, which are crucial in many real-world networks. To address these issues, we propose an information workflow $inc$-$LPCD_{AG}$, an incremental approach that predicts links and then detects communities in a dynamic environment. It consists of two phases: Phase-I predicts links to the attributed graph, and Phase-II forms communities by considering both the previous community-based and structural information of the updated graph. Therefore, with the proposed $inc$-$LPCD_{AG}$ information workflow, the Attributed Graph ($AG$) evolves gradually as new links are predicted and added, reflecting the dynamic changes of the $AG$ where the relationships and structure of the communities evolve.
\begin{figure} 
\center \includegraphics[width=4cm, height=6cm]{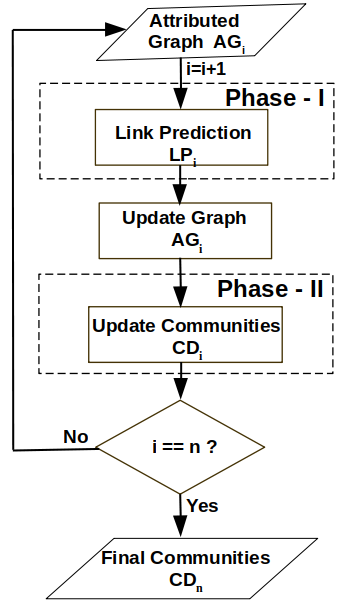}
\caption{Proposed information Workflow for Link Prediction and Community Detection in Dynamic Attributed Graphs ($inc$-$LPCD_{AG}$)}
\label{LPCD}
\end{figure}
Figure \ref{LPCD} represents the proposed $inc$-$LPCD_{AG}$ information workflow. For the given Attributed Graph ($AG_i$), where i=0, increment i and the Phase-I ($LP_i$) is executed for predicting the links based on structural and node attribute information of the graph or updated graph ($AG_i$). The predicted links are then added to the graph ($AG_i$). Further, the Phase-II will execute to form communities dynamically. Based on the condition, the procedure is repeated recursively, until $i=n$, at which point the final communities are obtained. For instance, in Figure \ref{LPCD} for a given attributed graph ($AG_{i=0}$), at the first iteration, `i' value is incremented, then the links are predicted ($LP_{i=1}$) and updated to the graph as ($AG_{i=1}$). Further, the communities ($CD_{i=1}$) are detected only using the structural information (including node attributes), as the previous community-based information is unavailable. From the second iteration, the procedure is repeated by executing two phases to obtain the final communities, $CD_n$ by evolving the relationships and structure of the communities till the maximum number of iterations is reached. Algorithm \ref{alg:LPCD_DAG} outlines the proposed $inc$-$LPCD_{AG}$ information workflow. 
\begin{table}\tiny
\centering
    \caption{Summary of Notations}
    \label{notations}
    \scalebox{0.8}{
    \begin{tabular}{c|l}
    \hline
    \textbf{Notation} & \textbf{\hspace{0.2cm}Description} \\
    \hline 
    $AG_{i}$ & \hspace{0.2cm}Attributed Graph at iteration `i' \\
    $n$ & \hspace{0.2cm}Number of iterations \\
    V & \hspace{0.2cm}Vertices (Nodes) \\
    E & \hspace{0.2cm}Edges (Links) \\
    A & \hspace{0.2cm}Node attributes \\
    $\mid E\mid$ & \hspace{0.2cm}Number of links in a Graph `G' \\
    $\mid V\mid$ & \hspace{0.2cm}Number of nodes in a Graph `G' \\
    $CD$ & \hspace{0.2cm}Communities \\
    \hline
    $LP$ & \hspace{0.2cm}Predicted links\\
    $V_{D}$ & \hspace{0.2cm}Affected nodes in dynamic graphs\\
    $L_{new}$ & \hspace{0.2cm}New links\\
    $d$ & \hspace{0.2cm}Distance between the nodes \\
    $Sim_{str}(v_i,v_j \mid d)$ & \hspace{0.2cm}Structural similarity between nodes $v_i$ and $v_j$ at distance $d$ \\
    $Sim_{attr}(v_i,v_j)$ & \hspace{0.2cm}Attribute similarity between nodes $v_i$ and $v_j$ \\
    $A_{ij}$ & \hspace{0.2cm}Value at $i^{th}$ row and $j^{th}$ column in Adjancency matrix $A$\\
    $\Gamma(v_i\mid d)$ & \hspace{0.2cm}Neighbouring nodes covered at distance $d$ for the node $v_i$\\
    $x_{v_i}$ & \hspace{0.2cm}Number of non-zero attributes at node $v_i$ \\
    $P(v_i\mid v_{i-1})$ & \hspace{0.2cm}Probability of having a node $v_{i}$ after the node $v_{i-1}$ in sequence \\
    $[W_{ij}]_{n\times n}$ & \hspace{0.2cm}Weighted matrix \\
    $[T_{ij}]_{n\times n}$ & \hspace{0.2cm}Transition matrix \\
    $\alpha$ & \hspace{0.2cm}Hyper-parameter to choose the portion of structural and node-attribute information \\
    $P\%$ & \hspace{0.2cm}Percentage of links to be added to the graph \\
    $RW$ & \hspace{0.2cm}Random walks\\
    $U$ & \hspace{0.2cm}Graph information with all connected and non-connected pairs of nodes\\
    $D$ & \hspace{0.2cm}Non-connected pairs of nodes\\
    $Walk\_length$ & \hspace{0.2cm}Length of the walk\\
    $Num\_Walks$ & \hspace{0.2cm}Number of walks in a sequence\\
    $N_E$ & \hspace{0.2cm}Node embeddings \\
    $E_E$ & \hspace{0.2cm}Edge embeddings for ($N_{E_i}$, $N_{E_j}$) \\
    \hline
    $Q_{prev}$ & \hspace{0.2cm}Modularity value of previous communities obtained \\
    $Q_{curr}$ & \hspace{0.2cm}Modularity value of current communities obtained \\
    $\Delta Q_{a \rightarrow C_2}$ & \hspace{0.2cm}Modularity gain for moving a node `a' to community $C_2$ \\
    $\Delta Q_{b \rightarrow C_1}$ & \hspace{0.2cm}Modularity gain for moving a node `b' to community $C_1$ \\
    $\Delta Q_{nei \rightarrow C_2}$ & \hspace{0.2cm}Modularity gain for moving a node `nei' to community $C_2$ \\
    $\Delta Q_{str}$ & \hspace{0.2cm}Modularity gain with respect to structural information\\
    $\Delta Q_{attr}$ & \hspace{0.2cm}Modularity gain for attribute information\\
    $k_{u,C_2}$ & \hspace{0.2cm}Number of nodes connected from node $u$ to the community $C_2$\\
    $\sum_{C_2}$ & \hspace{0.2cm}Sum of degrees of all the nodes in community $C_2$\\
    $k_u$ & \hspace{0.2cm}Degree of node `u'\\
    
    $\mu$ & \hspace{0.2cm}Center of gravity\\
    \textbf{X}$_n$ & \hspace{0.2cm}Sum of attribute values of each attribute type to the total number of nodes \\
    $I(V, v)$ & \hspace{0.2cm}{Inertia of `$V$' through `$v$'}\\
    $\left\| \mathbf{X}_v - \mu \right\|^2$ & \hspace{0.2cm}Euclidean distance from $\mathbf{X}_v$ to $\mu$ \\
    $I(V)$ & \hspace{0.2cm}{Inertia of nodes `$v \in C$' through its center of gravity} \\
    $\Delta Q$ & \hspace{0.2cm}{Modularity gain with respect to both structural and node attribute information}\\
    \hline
    {$\bm{|PL|}$} & \hspace{0.2cm}{Number of Predicted Links}\\
    {$\bm{|C^T|}$} & \hspace{0.2cm}{Number of Communities}\\
    \hline
    \end{tabular}}
\end{table}

\begin{algorithm}
\tiny
\SetAlgoLined
\KwIn{Attribute Graph $AG = (V, E, A)$} 
\KwOut{Communities $CD$}

\For{$Iteration\ i \leftarrow 1$ \KwTo $n$}{
    ${LP}_{i} \leftarrow \mathrm{DynamicCSADW}(AG_{i-1})$\;
    $AG_i \leftarrow AG_{i-1} \cup {LP}_{i}$\;
    $CD_i \leftarrow \mathrm{DynamicI\textnormal{-}Louvain}(AG_{i}, CD_{i-1}, LP_{i})$ \;
}
\Return{$CD_i$}
\caption{$inc$-$LPCD_{AG}$ Algorithm}
\label{alg:LPCD_DAG}
\end{algorithm}

\begin{figure}[!t]
\centering

\subfloat[Steps for the First Iteration\label{Exam1}]{
    \includegraphics[height=1.8cm,width=7cm]{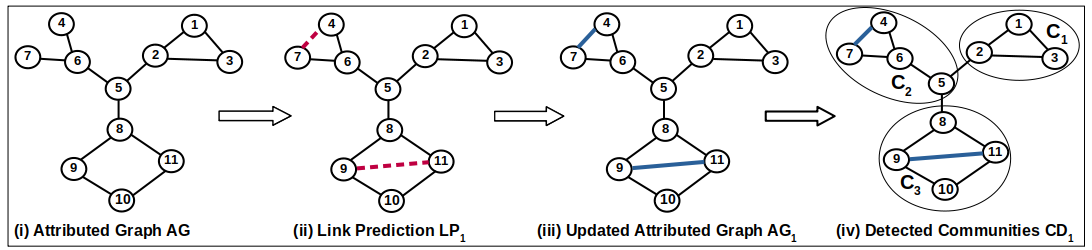}
}

\vspace{1em}

\subfloat[Steps for 2 to n Iterations\label{Exam2}]{
    \includegraphics[height=3.8cm,width=7cm]{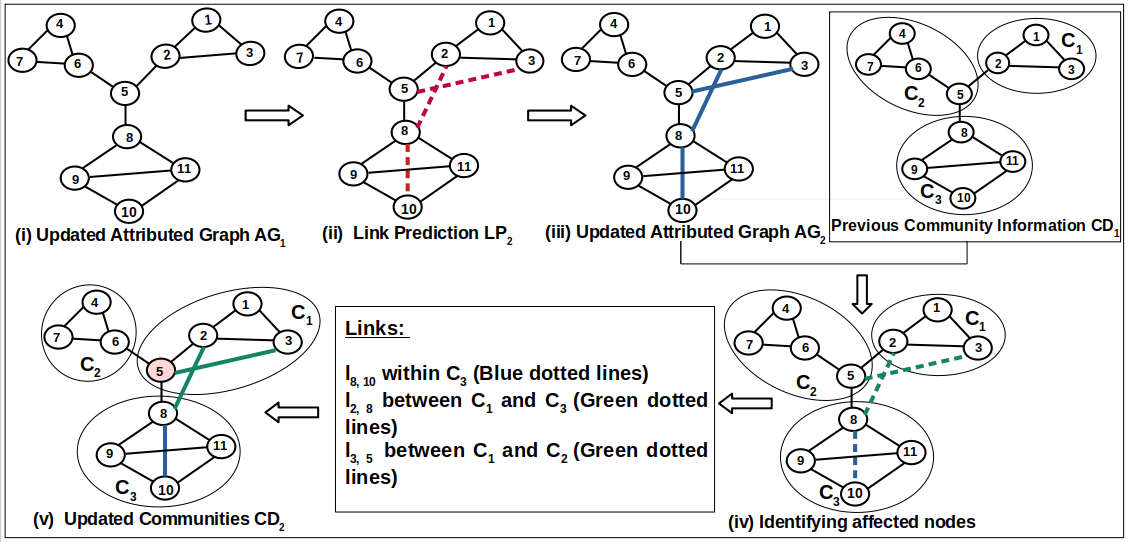}
}

\caption{Procedure for approach}
\label{Example}
\end{figure}

Let us consider an example shown in Figure \ref{Example}. The attributed graph consists of 11 nodes with 12 links, as shown in Figure \ref{Exam1}(i). At the first iteration, the links (4, 7) and (9, 11) are predicted, as shown in the dotted lines of Figure \ref{Exam1}(ii). Subsequently, the graph updates with these links in Figure \ref{Exam1}(iii). As there are no previous communities at the first iteration, the communities are detected for the updated graph of Figure \ref{Exam1}(iii) and three communities [$C_1$\{1, 2, 3\}, $C_2$\{4, 5, 6, 7\} and $C_3$\{8, 9, 10, 11\}] are formed as shown in Figure \ref{Exam1}(iv). For the second iteration, the updated graph, shown in Figure \ref{Exam2}(i) (same as Figure \ref{Exam1}(iii)) is considered, for which the links (2, 8), (3, 5), and (8, 10) are predicted as shown with dotted lines in Figure \ref{Exam2}(ii). Further, those links are added to the graph and shown in Figure \ref{Exam2}(iii). Then, the communities are detected using the updated graph (Figure \ref{Exam2}) and the previous community structure (Figure \ref{Exam1}(iv)) information by identifying the new links, whether they are within or between the communities (Figure \ref{Exam2}(iv)). According to \cite{DeltaScreening}, the community structure may change when the graph evolves with new links. If new links are within communities, there is no change in the community structure, but this might not be true when they are between communities. Therefore, the link (8, 10) is within the community, and the community structure will not change. Whereas, links (2, 8) and (3, 5) are between the communities, and the node will move to the desired community based on the modularity value between the nodes and their respective communities. For nodes 2, 8, and 3, there is no improvement in the modularity value and they remains in the same community whereas node 5 moved to $C_1$, as shown in Figure \ref{Exam2}(v), and $C_1$\{1, 2, 3, 5\}, $C_2$\{4, 6, 7\} and $C_3$\{8, 9, 10, 11\} are the final communities.  
\subsection{Predicting Links in Evolving Attributed Graphs: A Data-Driven Approach}\label{LPsubsection}
The Phase-I is for link prediction in the proposed $inc$-$LPCD_{AG}$ information workflow, and the details are provided in Algorithm \ref{alg:LPCD_DAG}. We have explored several approaches \cite{CSADW} \cite{PaGNN} \cite{RGNMF-AN} \cite{CPAGCN} for link prediction in attributed graphs, and found the CSADW \cite{CSADW} approach to be the most adaptable for link prediction (Phase-I) in the proposed information workflow. It is an embedding method that relies on the DeepWalk strategy, considering both the network structure and node attribute information of the attributed graph. This method is effective and designed to predict the links in a static environment. When applied to dynamic graphs, the node and edge embeddings are computed for the entire graph, which is time-consuming. To address this limitation, we proposed the Dynamic CSADW approach to support a dynamic environment, where the node and edge embeddings are incrementally updated whenever new links are added to the graph. These updated embeddings are then used to predict future links more efficiently.

Algorithm \ref{DynamicCSADWModified} outlines the procedure for Dynamic link prediction. Initially, generate node embeddings `$N_E$' for all nodes ($V_D$) in the graph (lines 2-4). If the graph is evolving with new predicted links ($L_{new}$), only affected node embeddings are computed and updated (lines 5-10). Then, compute the structural ($Sim_{str}$) and attribute similarity ($Sim_{attr}$) between each node in $V_D$ and its neighbours of the `$AG$' using (lines 11-14), and compute the weight matrix w($v_i$, $v_j$) using (lines 15-17). The transition matrix $[T_{ij}]_{n\times n}$ is computed from $[W_{ij}]_{n\times n}$, which is the probability of transforming to the next node (lines 18-20). Using $[T_{ij}]_{n\times n}$, the sequences of random walks are generated for each affected node ($V_D$) by deriving the predefined number of random walks ($num\_walks$) of length ($walk\_length$) and stored in $RW$ (lines 21-33). Further, $RW$ is an input to the skip-gram \cite{skipgram}, which produces the node embeddings and stores them in `$N_E$' (line 34).

To generate edge embeddings `$E_E$', initially, as $L_{new}$ is empty, consider the possible non-connected node pairs ($D$) (lines 35-36) as new links (lines 37-38), and generate edge embeddings ($E_E$) (lines 42-47). Then, these edge embeddings ($E_E$) are used as an input to the Logistic regression with L2 regularization \cite{logisticRegressionL2} to exhibit respective probability scores (lines 48-49). Based on these scores, according to the LINE \cite{LINE} approach, consider the first $P$ (3\%, 5\%, 10\%) links as predicted links ($L_{new}$) (lines 50-51). When this procedure adapts to the dynamic environment, evolving with new predicted links ($L_{new}$), edge embeddings ($E_E$) are excluded for $L_{new}$ (lines 39-41) and only for the affected links through $L_{new}$ are computed from the set of ($E_E$) and updated (lines 42-47) without recalculating from scratch.
\begin{algorithm}
\tiny
\setstretch{0.9}
\DontPrintSemicolon
\textbf{Procedure} \textproc{DynamicCSADW}($AG$)\;
\SetAlgoLined
\If{\textbf{not} $L_{new}$}{
	$V_{D} = AG.nodes()$
}
\Else{
\For{eachLink (u, v) in  $L_{new}$}{
	$V_{D}.append(u)$\;
	$V_{D}.append(v)$\;

}
}
\For{eachNode i $\in V_{D}$}{
\For{eachNei j $\in$ $AG.neighbor(i)$}{
$Sim_{str}(v_i,v_j\mid d) = A_{ij} \cdot \frac{\mid \Gamma(v_i\mid d)\cap\Gamma(v_j\mid d) \mid}{\mid \Gamma(v_i\mid d)\cup\Gamma(v_j\mid d)\mid}$\;
$Sim_{attr}(v_i,v_j) = \frac{\mid x_{v_i}\cap x_{v_j}\mid}{\mid x_{v_i}\cup x_{v_j}\mid}$\;
$w(v_i, v_j) = \alpha \times Sim_{attr} + (1-\alpha) \times Sim_{str}$\;
}
}
\For{eachNode i $\in$ $V_{D}$}{ 
	$[T_{ij}]_{n\times n} = \frac{[W_{ij}]_{n\times n}}{\sum_{j\in\Gamma(i)}[W_{ij}]_{n\times n}}$\; 
} 
\For{eachNode n $\in$ $V_{D}$}{
$RW[n] \leftarrow [\ ]$\;
\For{$w=1$ to $Num\_Walks$}{
	$walk \leftarrow [n]$ \;
        $current \leftarrow n$ \;
	
	\For{$wl = 1$ to $Walk\_Length$}{
		$NextNeiNode \leftarrow getNei(current, [T_{ij}]_{n\times n}) $\;
		$walk.append(NextNeiNode)$ \;
            	$current \leftarrow NextNeiNode$ \;
	}
	$RW[n].append(walk)$ \;	
}}
\BlankLine
$N_E \leftarrow$ skip-gram(RW)\; 
\If{\textbf{not} $L_{new}$}{
$D = U-AG.edges()$\;
$L_{new} = D$
}
\Else{
$D = D - L_{new}$ \;
}

\For{each (i, j) $\in L_{new}$ }{
	$k = D.getElement[i]$ \;
	$E_E[i, k] = Hadamard(N_{E_i}, N_{E_k})$\;
	$k = D.getElement[j]$ \;
	$E_E[j, k] = Hadamard(N_{E_j}, N_{E_k})$\;
	
}
$S = LogisticRegression.predict\_proba(E_{E})$\;
$Links \leftarrow$ SortDescending($S$)\;
$LP \leftarrow Links[:P \% AG.edges()]$\;
\Return{$LP$}
\caption{}
\label{DynamicCSADWModified}
\end{algorithm}

\subsection{Community Detection in Dynamic Attributed Graphs: A Knowledge-Driven Approach}\label{CDsubsection}
Phase-II is for Community Detection in the proposed $inc$-$LPCD_{AG}$ information workflow. We have explored a few techniques \cite{ILouvain} \cite{incAGGMMR} \cite{ModifiedLPA} for detecting communities in attributed graphs and adopted I-Louvain \cite{ILouvain}, a variant of the Louvain \cite{Louvain} algorithm. The reason for choosing I-Louvain is that, in one of our previous works \cite{LPCD}, we utilized the Louvain algorithm \cite{Louvain} for non-attributed graphs due to its proven efficiency. Given the effectiveness of the algorithm \cite{Louvain}, we extended the I-Louvain \cite{ILouvain} algorithm to operate in a dynamic environment and proposed the enhanced version, as the Dynamic I-Louvain algorithm. The I-Louvain \cite{ILouvain} algorithm effectively identifies communities by leveraging both node attributes and the structural information of the graph. It employs modularity maximization as its objective function, integrating inertia-based modularity to incorporate node attribute information. The Dynamic I-Louvain algorithm in Phase-II of the proposed information workflow handles dynamic updates without reconstructing the communities from scratch. This algorithm leverages both the information 1) Predicted links and 2) Existing community structure (previously detected communities) to ensure efficiency and to reduce cost without processing from scratch.

In Dynamic I-Louvain, when new links are added to `$AG$', only the nodes affected by those links are considered without performing the process on the unaffected nodes. For these affected nodes, communities are updated incrementally, based on whether links are added within or between communities in an attributed graph. If links are added within the communities, the community structure remains the same. If the links are added between the communities, the affected nodes and their corresponding communities are updated. This strategy, which adheres to the principles of Zarayeneh et al. \cite{DeltaScreening} for non-attributed graphs, is efficient. Hence, we adopted the same principle for attributed graphs and proposed the Dynamic I-Louvain algorithm.
\begin{algorithm}
\tiny
\setstretch{0.9}
\caption{} 
\label{ILouvain_new}
\DontPrintSemicolon
\textbf{Procedure} \textproc{DynamicI-Louvain}($AG$, $CD$, $LP$)\;
\SetAlgoLined
\If{$CD == \phi$}{
	\For{$each\ n \in AG.nodes()$}{
		$CD[n] = n$\;
	}
	$LP = AG.edges()$\;
	$Q_{prev}= Q_{curr} = 0$\;
	$converge = False$\;
}
\SetKwBlock{Repeat}{Repeat}{Until{ $\mathit{converge == {True}}$}}
\Repeat{
\For{\textbf{each} $(a, b) \in LP$}{
	
	$C_1 \leftarrow \{C_i \mid a \in CD[i]\}$\;
	$C_2 \leftarrow \{C_i \mid b \in CD[i]\}$\;
	\If{$C_1 == C_2$}{
		$Continue$\;
	}
	$\Delta Q_{a\rightarrow C_2} = ComputeModularityGain(a, C_1, C_2)$\;
	$\Delta Q_{b\rightarrow C_1} = ComputeModularityGain(b, C_2, C_1)$\;
	\If{$\Delta Q_{a\rightarrow C_2} \ge \Delta Q_{b\rightarrow C_1}\ \mathbf{and}\ \Delta Q_{a\rightarrow C_2}>0 $}{
		$C_2.append(a)$\;
		$C_1.remove(a)$\;
		\For{\textbf{each} nei $\in$ $AG.neigbours(a) \ \mathbf{and} \ nei \in C_1$}{
			$\Delta Q_{nei\rightarrow C_2} = ComputeModularityGain(nei, C_1, C_2)$\;
			\If{$\Delta Q_{nei\rightarrow C_2} > 0$}{
				$C_2.append(nei)$\;
				$C_1.remove(nei)$\;
			}
		}
		
	}
	\ElseIf{$\Delta Q_{b\rightarrow C_1} > 0$}{
		$C_1.append(b)$\;
		$C_2.remove(b)$\;
		\For{\textbf{each} nei $\in$ $AG.neigbours(b) \ \mathbf{and} \ nei \in C_2$}{
			$\Delta Q_{nei\rightarrow C_1} = ComputeModularityGain(nei, C_2, C_1)$\;
			\If{$\Delta Q_{nei\rightarrow C_1} > 0$}{
				$C_1.append(nei)$\;
				$C_2.remove(nei)$\;
			}
		}
	}
	
}
$Q_{curr} = modularity(AG, CD)$\;
\If{$converge == False$}{
	\If{$Q_{prev} == Q_{curr}$}{
	$converge = True $\;
	$break$\;
	}
	$Q_{prev} = Q_{curr}$\;
	}
}
\Return{$CD$}
\end{algorithm}

Algorithm \ref{ILouvain_new} outlines the procedure for updating communities using the proposed Dynamic I-Louvain algorithm. $CD$ is the list of communities used to maintain community structure information, and it is initially empty. The process starts by considering each node as an individual community and storing it in $CD$ (lines 2-5). The variables $Q_{prev}$ and $Q_{curr}$ store previous and current modularity values, respectively, and are both initialized to zero (line 7). The boolean variable \textit{converge} is a flag and initialized to \textit{False} (lines 8-9). Each node in a node pair based on $LP$ information (lines 11-18) is evaluated by computing the modularity value (Algorithm \ref{ILouvainMod}) through a process of temporarily moving to other communities (lines 17-41). This process is repeated iteratively until the modularity value no longer changes, indicating that no node movement improves modularity (lines 42-49). At this point, the \textit{converge} flag is set to \textit{True} (lines 50) with the obtained set of communities $CD$ (line 51).

In the incremental steps, when the new links ($LP$) are predicted, the process aims to update community structures by identifying communities of node `$a$' and node `$b$' (say $C_1$ and $C_2$ respectively) (lines 10-13). If node `$a$' and node `$b$' lie in the same community, then continue (lines 15-16). Otherwise, compute the modularity gain (Algorithm \ref{ILouvainMod}) of moving a node `$a$' to $C_2$ ($\Delta Q_{a\rightarrow C_2}$) and node `$b$' to $C_1$ ($\Delta Q_{b\rightarrow C_1}$) (lines 17-18). When $\Delta Q_{a\rightarrow C_2}$ is positive and greater than or equal to $\Delta Q_{b\rightarrow C_1}$, then append node `$a$' to community $C_2$ and remove from community $C_1$ (lines 19-21). Further, for each neighbour node `$nei$' of node `$a$' with respect to $C_1$, compute the modularity of moving it to $C_2$. If the modularity (lines 22-24) is greater than 0, then append node `$nei$' to $C_2$ and remove it from $C_1$ (lines 25-29); otherwise, it remains in the same community. Similarly, repeat the same procedure for condition $\Delta Q_{b\rightarrow C_1}$ greater than 0 (lines 30-41) and for neighbouring nodes of node `$b$'. This process is repeated for all predicted links $LP$, and the final communities are obtained after the subsequent `$n$' incremental steps (lines 51).
\begin{algorithm}
\tiny
\caption{} 
\label{ILouvainMod}
\DontPrintSemicolon
\textbf{Procedure} \textproc{ComputeModularityGain}($u$, $C_1$, $C_2$)\;
\SetAlgoLined
$\Delta Q_{str} = \frac{k_{u,C_2}}{2 \mid E\mid} - \frac{k_u \sum{k_{C_2}}}{2|E|^2}$\;
$\mu = \frac{1}{|V|} \sum_{n \in V} \mathbf{X}_n$\;
$I(V, v) = \left\| \mathbf{X}_v - \mu \right\|$\;
$I(V) = \sum_{v \in V} \left\| \mathbf{X}_v - \mu \right\|^2$\;
$A = \sum_{v\in C_2 } \left[ \frac{I(V, u) \cdot I(V,v)}{2\mid V\mid \cdot I(V)} - {\left\lVert u, v \right\rVert}^{2} \right]$\;
$B = \sum_{v\in C_1} \left[ \frac{I(V,u) \cdot I(V, v)}{2\mid V\mid \cdot I(V)}-{\left\lVert u,v \right\rVert} ^{2} \right] $\;
$\Delta Q_{attr} = \frac{1}{\mid V\mid \cdot I(V)} (A-B) $\;
$\Delta Q = \Delta Q_{str} + \Delta Q_{attr}$\;
\Return{$\Delta Q$}
\end{algorithm}
\subsection{Computational Complexity Analysis}

The time complexity of Phase-I: $O(V^2 + V\times Num\_Walks \times Walk\_Length + M + E\times d)$, where computing the structural and node attribute similarity, weight matrix and transition matrix takes $O(V\times(V-1))$. Then, to generate the sequences of random walks, the time taken is $O(V\times Num_Walks \times Walk\_Length)$, as for each node in $V_D$, the number of random walks ($Num\_Walks$) is generated of size ($Walk\_Length$). Then, to generate node embeddings using the skip-gram model, assuming the time taken is $O(M)$. To generate the non-connected pairs the time taken is $O(V^2)$. The edge embeddings are generated with the Hadamard operator with time complexity $O(E\times d)$.  

The time complexity of Phase-II: $O(k \times E \times (V-1) + E\times [(V-1) + (V-1)])$ which is approximately $O(E\times (V-1))$. For each new link, computing modularity for a nodes along the links and its neighbour takes $O(E \times (V-1) \times (V-1))$  time. If the convergence is after the $k$ iterations then the time taken is $O(k\times E\times (V-1)) = O(E\times V)$

The overall time complexity of $inc$-$LPCD_{AG}$ information workflow is $O(n \times [(V^2 + V\times Num\_Walks \times Walk\_Length + M + E\times d) + (E\times (V-1))])$ where $n$ is number of incremental steps.

\section{Experimental setup, Datasets}
We conducted experiments on our proposed information workflow using the benchmark dataset presented in Section \ref{dataset}. All experiments were executed on a desktop equipped with an Intel i7 1.90 GHz processor and 32 GB of RAM on an Ubuntu 20.04.3 LTS operating system with Python version 3.9.23. The proposed information workflow, $inc$-$LPCD_{AG}$, consists of two phases: Link Prediction and Community Detection. We employed benchmark datasets suitable for both phases and experimented to validate our proposed information workflow. Phase-I addresses the link prediction task and is detailed in Section \ref{LPsubsection}, while Phase II focuses on community detection, as described in Section \ref{CDsubsection}. To evaluate Phase-I performance, the AUC and AP metrics \cite{CSADW} are used. And for Phase-II, Performance \cite{ASMsg}, Modularity \cite{DSLPA}, Density \cite{CEMOV}, and Conductance \cite{FLCDA} are used.

\subsection{Computational Environment}
The proposed $inc$-$LPCD_{AG}$ information workflow will predict the links and detect communities in a dynamic environment. As considering all non-connected pairs for link prediction results in exponential time, we have conducted experiments by considering non-connected pairs at a 2-hop distance \cite{twohop}. We have performed experiments over different sets of iterations, such as \{5, 10, 15, 20\}, to predict links. Therefore, the top-ordered links are derived from the list of non-connected pairs (3\%, 5\%, and 10\% of connected edges in the graph) and treated as predicted links to update the graph, rather than considering the entire list.
\subsection{Benchmark Datasets}\label{dataset}
The benchmark datasets used for our information workflow are Cora \cite{coraapp}, Citeseer \cite{citeseerapp}, and the Twitter \cite{twitter} dataset. Cora \cite{coraapp} and Citeseer \cite{citeseerapp} are the bibliographic datasets where the nodes represent a scientific publication, and links represent the citation links between the publications. The Cora dataset consists of 2708 nodes, each node is described by 1433 binary attributes and having 5429 links whereas Citeseer dataset consists of 3312 nodes, each node has 3703 binary attributes and having 4732 links. 

Twitter \cite{twitter} dataset consists of 2511 nodes with one node attribute \textit{Tag} and 37154 edges. In one of our works $Synt\_H_oAG$ approach in $SDG\_HHAG$ \cite{Synth} framework is designed to generate synthetic dataset by enriching with the node and edge attributes. As this dataset consists of only one node attribute, \textit{Tag}, we have enriched the node attributes with additional four attributes: \textit{Visibility}, \textit{Gender}, \textit{Age}, and \textit{Followers}, with our $Synt\_H_oAG$ approach and performed experiments and named it as $Twitter^+$ dataset.

\section{Results, Validation and Statistical Analysis}
This section presents the results of the proposed $inc$-$LPCD_{AG}$ approach and evaluates its performance against conventional approaches. The proposed approach comprises two phases: link prediction in Phase-I, which is independently evaluated, and the intermediate results are presented in Table \ref{LPEval1}. The final outcome of Phase-II is shown in Table \ref{CommunityDetectionMeasure}. In addition, we have performed statistical analysis by showing that our proposed approach is effective and statistically significant. 
\subsection{Significance of Results}
The proposed $inc$-$LPCD_{AG}$ approach is applied to various benchmark datasets, and the corresponding Phase-I and Phase-II results are presented in the \textbf{F} and \textbf{G} columns of Table \ref{CommunityDetectionMeasure}. Further, the final communities are evaluated through Performance \cite{ASMsg}, Modularity \cite{DSLPA}, Density \cite{CEMOV}, and Conductance \cite{FLCDA}.  To demonstrate the dynamic environment of the proposed approach, we have conducted experiments over 20 iterations \{5, 10, 15, 20\} to simulate incremental updates at each iteration, and the results are recorded for every 5 iterations as shown in column \textbf{C}. The percentages (3\%, 5\%, and 10\%) of links are predicted and updated in the graph to obtain final communities. The results are evaluated with benchmark measures and provided in columns \textbf{H}, \textbf{I}, \textbf{J}, and \textbf{K}. It is observed that while 3\% and 5\% links are predicted, the quality of communities is good over the iterations. For 10\% of links, the values of the measures decrease slightly while retaining community quality, because of the increase in number of predictions, particularly when links are predicted between communities, which certainly reduces community quality. For the $Twitter^+$ dataset with 10\% of links, results are provided up to the $15^{th}$ iteration, as at the $17^{th}$ iteration, no sufficient non-connected pairs remain under the two-hop distance constraint.
\begin{table}
\centering
\renewcommand{\arraystretch}{0.8}
\caption{\scriptsize{Evaluating the Proposed information Workflow $inc$-$LPCD_{AG}$}}
    \label{CommunityDetectionMeasure}
    \scalebox{0.25}{
    \Huge \begin{tabular}{|l|c|c|c|c|c|c|c|c|c|c|}
    \hline
    \multirow{2}{*}{\textbf{Dataset}} & \multirow{2}{*}{\textbf{D\% links}} & \multirow{2}{*}{\textbf{\#Iterations}} & \multirow{2}{*}{$\bm{|E|}$} & \multirow{2}{*}{$\bm{|E^-|}$} & \textbf{Phase-I} & \textbf{Phase-II}  &\multicolumn{4}{c|}{\textbf{Measures}} \\
        \cline{6-11} 
        &  &  & & & {$\bm{|PL|}$} & {$\bm{|C^{T}|}$}  & {\textbf{P} ($\uparrow$)} & {\textbf{Q} ($\uparrow$)} & {$\bm{\rho}$ ($\uparrow$)} & {$\bm{\phi}$ ($\downarrow$)} \\
        \hline
        \hspace{1cm}\textbf{A} & \textbf{B} & \textbf{C} & \textbf{D} & \textbf{E} & \textbf{F} & \textbf{G} & \textbf{H} & \textbf{I} & \textbf{J} & \textbf{K} \\
        \hline
        \multirow{12}{*}{Cora} & \multirow{4}{*}{3} & 5 & 6116 & 498442 & 178 & 99 & 0.678 & 0.604 & 0.842 & 0.019 \\ 
        \cline{3-11}
         & & 10 & 7087 & 498264 & 206 & 98 & 0.683 & 0.610 & 0.852 & 0.017 \\
        \cline{3-11}
         & & 15 & 8213 & 498058 & 239 & 98 & 0.687 & 0.615 & 0.859 & 0.017 \\
        \cline{3-11}
        & & 20 & 9519 & 497819 & 277 & 98 & 0.687 & 0.617 & 0.866 & 0.017 \\
        \cline{2-11}
        \noalign{\vskip 2pt}
	\cline{2-11}
         & \multirow{4}{*}{5} & 5 & 6733 & 498442 & 320 & 99 & 0.744 & 0.666 & 0.832 & 0.020 \\
        \cline{3-11}
         & & 10 & 8591 & 498122 & 409 & 99 & 0.748 & 0.671 & 0.850 & 0.019 \\
        \cline{3-11}
         & & 15 & 10963 & 497713 & 522 & 96 & 0.749 & 0.673 & 0.863 & 0.020 \\
        \cline{3-11}
         & & 20 & 13990 & 497191 & 666 & 99 & 0.746 & 0.673 & 0.877 & 0.021 \\
	\cline{2-11}
	\noalign{\vskip 2pt}
	\cline{2-11}
	 & \multirow{4}{*}{10} & 5 & 8497 & 498442 & 772 & 99 & 0.629 & 0.580 & 0.868 & 0.017 \\
	\cline{3-11}
	 & & 10 & 13681 & 497670 & 1243 & 92 & 0.637 & 0.588 & 0.887 & 0.019 \\
	\cline{3-11}
	 & & 15 & 22030 & 496427 & 2002 & 95 & 0.621 & 0.570 & 0.909 & 0.022 \\
	\cline{3-11}
	 & & 20 & 35478 & 494425 & 3225 & 88 & 0.614 & 0.560 & 0.927 & 0.018 \\
	\hline
	\noalign{\vskip 2.2pt}
	\hline
        \multirow{12}{*}{Citeseer} & \multirow{4}{*}{3} & 5 & 5257 & 153456 & 153 & 409 & 0.943 & 0.875 & 0.886 & 0.003 \\ 
        \cline{3-11}
         & & 10 & 6092 & 153303 & 177 & 409 & 0.943 & 0.879 & 0.907 & 0.003 \\
        \cline{3-11}
         & & 15 & 7059 & 153126 & 205 & 409 & 0.944 & 0.880 & 0.920 & 0.003 \\
        \cline{3-11} 
         & & 20 & 8181 & 152921 & 238 & 409 & 0.944 & 0.878 & 0.931 & 0.003 \\
        \cline{2-11}
        \noalign{\vskip 2.2pt}
	\cline{2-11}
         & \multirow{4}{*}{5} & 5 & 5787 & 153456 & 275 & 412 & 0.934 & 0.872 & 0.908 & 0.002 \\
        \cline{3-11}
         & & 10 & 7382 & 153181 & 351 & 412 & 0.935 & 0.874 & 0.928 & 0.002 \\
        \cline{3-11}
         & & 15 & 9419 & 152830 & 448 & 412 & 0.935 & 0.865 & 0.940 & 0.002 \\
        \cline{3-11}
         & & 20 & 12019 & 152382 & 572 & 403 & 0.937 & 0.855 & 0.947 & 0.003 \\
	\cline{2-11}
	\noalign{\vskip 2.2pt}
	\cline{2-11}
	& \multirow{4}{*}{10} & 5 & 7301 & 153456 & 663 & 409 & 0.940 & 0.875 & 0.923 & 0.003 \\
	\cline{3-11}
	 & & 10 & 11756 & 152793 & 1068 & 407 & 0.942 & 0.860 & 0.942 & 0.004 \\
	\cline{3-11}
	 & & 15 & 18931 & 151725 & 1721 & 405 & 0.945 & 0.833 & 0.952 & 0.005 \\
	\cline{3-11}
	& & 20 & 30486 & 150004 & 2771 & 405 & 0.947 & 0.799 & 0.957 & 0.006 \\
	\hline
	\noalign{\vskip 2.2pt}
	\hline
        \multirow{12}{*}{Twitter$^+$} & \multirow{4}{*}{3} & 5 & 43069 & 127454 & 1292 & 22 & 0.926 & 0.872 & 0.508 & 0.041 \\ 
        \cline{3-11}
         & & 10 & 48472 & 123343 & 1454 & 22 & 0.929 & 0.876 & 0.542 & 0.037 \\
        \cline{3-11}
         & & 15 & 56189 & 115626 & 1685 & 22 & 0.932 & 0.880 & 0.589 & 0.039 \\
        \cline{3-11}
         & & 20 & 65136 & 106679 & 1954 & 22 & 0.936 & 0.881 & 0.634 & 0.040 \\
        \cline{2-11} 
        \noalign{\vskip 1.8pt}
        \cline{2-11}
         & \multirow{4}{*}{5} & 5 & 45159 & 126656 & 2257 & 23 & 0.93 & 0.865 & 0.542 & 0.050  \\
        \cline{3-11}
         & & 10 & 57632 & 114183 & 2881 & 19 & 0.745 & 0.840 & 0.646 & 0.170   \\
        \cline{3-11}
         & & 15 & 73551 & 98264 & 3677 & 18 & 0.754 & 0.850 & 0.801 & 0.134   \\
        \cline{3-11}
         & & 20 & 93870 & 77945 & 4693 & 17 & 0.771 & 0.843 & 0.890 & 0.087 \\
	\cline{2-11}
	\noalign{\vskip 2.3pt}
	\cline{2-11}
	 & \multirow{4}{*}{10} & 5 & 54395 & 111981 & 5439 & 22 &  0.936 & 0.872 & 0.574 & 0.037   \\
	\cline{3-11}
	 & & 10 & 87600 & 75455 & 8760 & 21 & 0.931 & 0.868 & 0.767 & 0.041   \\
	\cline{3-11}
	& & 15 & 141079 & 16629 & 14107 & 18 & 0.911 & 0.778 & 0.816 & 0.131  \\
        \hline
        \multicolumn{10}{l}{\# - Number of }
        
    \end{tabular}}
\end{table}

\subsubsection{Parameter Tuning}
The proposed $inc$-$LPCD_{AG}$ approach, especially Phase-I, involves parameters such as $\alpha$, $Walk\_Length$, and $Num\_Walks$. We systematically explored various values through experiments to identify optimal settings.
\begin{itemize}
\item`$\alpha$' controls the balance between structural information and node attribute information. We experimented with the range 0.4 to 0.7 and found that promising results are obtained at 0.6, which indicates a focus on structural importance rather than node attribute information. 
\item `$Walk\_Length$' determines the steps in a random walk. Testing values from 20 to 90 showed that 80 steps effectively capture node relationships, highlighting the optimal choice.  
\item `$Num\_Walks$' is the multiple random walks from the same node, and the range 5 to 25 is considered; it is found that 20 walks per node is effective in retrieving the neighbouring node information.
\item `node embeddings' are generated based on the size of the vector, and conducted experiments with 24 to 28, found that the vector size 128 is suitable for all datasets.  
\end{itemize}

\subsection{Evaluating the proposed approach}
We have evaluated the proposed approach, considering attributed graphs and relaxing node attributes to treat them as non-attributed graphs, and compared it with conventional methods to demonstrate its suitability for all circumstances and applicability across various domains. For comparison purposes, we have considered 10\% of predicted links over 20 iterations for Cora and Citeseer datasets, whereas 15 iterations for the Twitter$^+$ dataset, and the results are shown in Tables \ref{incAGGMMRvsincLPCD} and \ref{LPCDvsincLPCD}.
\subsubsection{Performance Comparison on Attributed Graphs}
The proposed $inc$-$LPCD_{AG}$ approach is compared with the conventional approach, $inc$-$AGGMMR$ \cite{incAGGMMR}, and shown in Table \ref{incAGGMMRvsincLPCD}. We have experimented by adding edges to the $inc$-$AGGMMR$ \cite{incAGGMMR} to compare with our proposed approach. It is observed that the results of the evaluation measures- modularity ($Q$), density ($\rho$), and conductance ($\phi$) - show that $inc$-$LPCD_{AG}$ outperforms the conventional method, as reflected in columns \textbf{F}, \textbf{G}, \textbf{H}, and \textbf{I}

\begin{table}
\centering
\caption{\scriptsize{Comparison of Proposed $inc$-$LPCD_{AG}$ information Workflow with existing inc-AGGMMR \cite{incAGGMMR} approach}}
    \label{incAGGMMRvsincLPCD}

	\scalebox{0.5}{
  \begin{tabular}{|c||c|c|c|c||c|c|c|c|}
     \hline
    	\multirow{3}{*}{\textbf{Dataset}} & \multicolumn{8}{c|}{\textbf{Evaluation of Communities}} \\
    	 \cline{2-9} 
         &  \multicolumn{4}{c||}{{$\bm{inc}-\bm{AGGMMR}$ \textbf{\cite{incAGGMMR}}}} &  \multicolumn{4}{c|}{{$\bm{inc}$\textbf{-}$\bm{LPCD_{AG}}$}} \\
        \cline{2-9} 
        & \textbf{P} ($\uparrow$) & \textbf{Q} ($\uparrow$) & \textbf{$\bm{\rho}$} ($\uparrow$) & \textbf{$\bm{\phi}$} ($\downarrow$) & \textbf{P} ($\uparrow$) & \textbf{Q} ($\uparrow$) & \textbf{$\bm{\rho}$} ($\uparrow$) & \textbf{$\bm{\phi}$} ($\downarrow$) \\
        \hline
        \hspace{0.8cm}\textbf{A} & \textbf{B} & \textbf{C} & \textbf{D} & \textbf{E} & \textbf{F} & \textbf{G} & \textbf{H} &  \textbf{I}\\
	\hline
	{Cora} & \textbf{0.943} & 0.119 & 0.033 & 0.830 & 0.614 & \textbf{0.560} & \textbf{0.927} & \textbf{0.018} \\
	\hline
	
	{Citeseer} & 0.928 & 0.161 & 0.019 & 0.769 & \textbf{0.947} & \textbf{0.799} & \textbf{0.957} & \textbf{0.006} \\
	\hline
	{Twitter$^+$} & 0.864 & 0.178 & 0.136 & 0.713 & \textbf{0.911} & \textbf{0.778} & \textbf{0.816} & \textbf{0.131} \\
	\hline
    \end{tabular}}
\end{table}
\subsubsection{Performance Comparison on Non-Attributed Graphs}
The proposed $inc$-$LPCD_{AG}$ approach is compared with the conventional LPCD \cite{LPCD} method on non-attributed graphs, and shown in Table \ref{LPCDvsincLPCD}. To demonstrate its versatility, experiments are conducted with the proposed $inc$-$LPCD_{AG}$ by relaxing the attribute information (termed as $inc$-$LPCD$) and comparing the results with those from one of our earlier works, LPCD \cite{LPCD}. The results show that $inc$-$LPCD_{AG}$ performs better, especially with the modularity measure, and this is evident in the quality of the final communities. It is also observed that with other measures, the proposed approach shows comparable performance across datasets.
\begin{table}
\centering
\caption{\scriptsize{Comparison between Conventional and Proposed approaches}}
    \label{LPCDvsincLPCD}

	\scalebox{0.5}{
  \begin{tabular}{|l||c|c|c|c||c|c|c|c|}
     \hline
    	\multirow{3}{*}{\textbf{Dataset}} & \multicolumn{8}{c|}{\textbf{Evaluation of Communities}} \\
    	 \cline{2-9} 
         &  \multicolumn{4}{c||}{{$\bm{LPCD}$ \textbf{\cite{LPCD}}}} &  \multicolumn{4}{c|}{{$\bm{inc}$\textbf{-}$\bm{LPCD}$}} \\
        \cline{2-9} 
        & \textbf{P} ($\uparrow$) & \textbf{Q} ($\uparrow$) & \textbf{$\bm{\rho}$} ($\uparrow$) & \textbf{$\bm{\phi}$} ($\downarrow$) & \textbf{P} ($\uparrow$) & \textbf{Q} ($\uparrow$) & \textbf{$\bm{\rho}$} ($\uparrow$) & \textbf{$\bm{\phi}$} ($\downarrow$) \\
        \hline
        \hspace{0.8cm}\textbf{A} & \textbf{B} & \textbf{C} & \textbf{D} & \textbf{E} & \textbf{F} & \textbf{G} & \textbf{H} & \textbf{I} \\
        \hline
	{Cora} & 0.951 & 0.801 & 0.824 & \textbf{0.021} & \textbf{0.955} & \textbf{0.862} & \textbf{0.870} & \textbf{0.021} \\
	\hline
	
	{Citeseer} & 0.982 & 0.759 & 0.977 & \textbf{0.004} & \textbf{0.984} & \textbf{0.888} & \textbf{0.992} & 0.005 \\
	\hline
	{Twitter$^+$} &  \textbf{0.969} &  \textbf{0.888} & \textbf{0.873} & \textbf{0.013} & {0.943} & {0.729} & {0.794} & {0.131} \\
	\hline
    \end{tabular}}
\end{table}

\subsubsection{Ablation Study}
To evaluate the Link Prediction (Phase-I) of the proposed approach, we used the AUC \cite{CSADW} and AP \cite{CSADW} metrics, as shown in Table \ref{LPEval1}. To demonstrate and carry out the experiments, we have considered the graph at four iterations as the Iteration Graph ($IG_k$). Initially, the input attributed graph is divided into two parts: one part, i.e., subgraph ($SG_1$) for link prediction is considered at $IG_1$ and the other ($SG_2$) for evaluation as ground truth. This iterative process, in which correctly predicted links are inserted into $SG_1$ for subsequent predictions, highlights the method's thoroughness. The same procedure is followed for the remaining iterations, and the results are provided in Table \ref{LPEval1}. It is observed that the $AUC$ and $AP$ results are good, especially when the graph is updated with predicted links, indicating that our dynamic link prediction (Phase-I) performs better in a dynamic environment.  
\begin{table}[h!]
\centering
\caption{\scriptsize{Evaluating the Link Prediction}}
    \label{LPEval1}
    \scalebox{0.18}{
    \Huge 
    \begin{tabular}{|l|c|c|c|c|c||c|c|c|c|c||c|c|c|c|c||c|c|c|c|c||}
    \hline
    	\multirow{2}{*}{\textbf{Dataset}} & \multicolumn{5}{c||}{\textbf{$\bm{IG_1}$}} & \multicolumn{5}{c||}{\textbf{$\bm{IG_2}$}} & \multicolumn{5}{c||}{\textbf{$\bm{IG_3}$}} & \multicolumn{5}{c|}{\textbf{$\bm{IG_4}$}} \\
        \cline{2-21}        
        & $\bm{|E|}$ & $\bm{|PL|}$ & $\bm{|CPL|}$ & \textbf{AUC} & \textbf{AP} & $\bm{|E|}$ & $\bm{|PL|}$ & $\bm{|CPL|}$ & \textbf{AUC} & \textbf{AP} & $\bm{|E|}$ & $\bm{|PL|}$ & $\bm{|CPL|}$ & \textbf{AUC} & \textbf{AP} & $\bm{|E|}$ & $\bm{|PL|}$ & $\bm{|CPL|}$ & \textbf{AUC} & \textbf{AP} \\ 
        \hline
        Cora & 2484 & 640 & 391 & 0.674 & 0.737 & 2875 & 911 & 566 & 0.690 & 0.744 & 3441 & 985 & 637 & 0.702 & 0.764 & 4078 & 991 & 659 & 0.727 & 0.800 \\ 
        \hline
        Citeseer & 2109 & 386 & 260 & 0.775 & 0.825 & 2369 & 514 & 319 & 0.707 & 0.763 & 2688 & 580 & 450 & 0.840 & 0.863 & 3138 & 530 & 422 & 0.872 & 0.902 \\ 
	\hline
        Twitter$^+$ & 2005 & 6258 & 5244 & 0.883 & 0.886 & 7249 & 7272 & 5987 & 0.863 & 0.885 & 13236 & 7542 & 6082 & 0.854 & 0.874 & 19318 & 7729 & 6276 & 0.845 & 0.879 \\
        \hline
    \end{tabular}}
\end{table}

\subsection{Statistical Analysis}\label{statisticalanalysis}
To validate the effectiveness of the proposed $inc$-$LPCD_{AG}$ approach against the conventional approach, $inc$-$AGGMMR$ \cite{incAGGMMR}, we conducted a statistical significance analysis. These approaches were evaluated using \textit{Paired t-test} across various benchmark datasets. The test produced a p-value of 0.01142, indicating that the proposed $inc$-$LPCD_{AG}$ approach performs significantly better than the conventional approach inc-AGGMMR \cite{incAGGMMR}. Additionally, \textit{Cohen’s d} is used to quantify the magnitude of the difference between two approaches, using a set of Modularity values, and yielded a result of 5.8291, indicating a large effect size. This highlights the significance of the proposed $inc$-$LPCD_{AG}$ approach, which includes Phase-I as a link prediction strategy in the information workflow, thereby making the network structure cohesive and allowing community detection algorithms to form stable communities rather than randomly adding edges to simulate the dynamic environment in the inc-AGGMMR \cite{incAGGMMR}. The importance of link prediction is especially evident in sparse or incomplete networks.

\section{Conclusions and Future Work}
The dynamics of the social systems prevail and analysing them based on structural behaviour for several applications is challenging. Thus, an information workflow $inc$-$LPCD_{AG}$ in dynamic attributed graphs will fulfil to meet the requirements especially involved with real word graphs. This information workflow supports by analysing and understanding the needs involved with attributed graphs. Additionally, each phase is enhanced in the workflow to support the dynamic environment. We have evaluated our approach to demonstrate its efficiency and compared its results with those of existing methods. In the future, this information workflow can be extended to weighted attributed graphs, where edge weights provide additional information to help in evolved social systems in estimating the relationships. And also, with the cohesive properties, communities can be formed and will be beneficial for diverse applications. Also, our proposed approach has a scope to handle different types of nodes and its attributes rather single type of node.  
\section*{Acknowledgements}
We would like to acknowledge Dr. Arun Kumar Das for his constructive suggestions and thoughtful insights that substantially strengthened the computational complexity analysis. We further acknowledge our research team for their valuable feedback and continous support.

\bibliographystyle{plain}
\bibliography{ref} 
\end{document}